\pdfoutput=1
\documentclass[cernpreprint, texlive=2023, UKenglish, texmf, orcidlogo]{atlasdoc}
 
\usepackage{todonotes}
\DeclareBoolOption[true]{LANGEDIT}
\LEcontact{Chip Brock brockr@msu.edu}

\usepackage[subfigure]{atlaspackage}
\usepackage{atlasbiblatex}
 
\usepackage[BSM=true,jetetmiss=true]{atlasphysics}

\graphicspath{{logos/}{figures/}}
 
\usepackage{ANA-HDBS-2018-45-PAPER-defs}
 
\usepackage{multirow}

\AtlasTitle{Search for Resonant Production of Dark Quarks in the Dijet Final State with the ATLAS Detector}
 
\AtlasAbstract{
This paper presents a search for a new $Z^\prime$ resonance decaying into a pair of dark quarks which hadronise into dark hadrons before promptly decaying back as Standard Model particles. This analysis is based on proton--proton collision data recorded at $\sqrt{s}=13$ TeV with the ATLAS detector at the Large Hadron Collider between 2015 and 2018, corresponding to an integrated luminosity of 139 fb$^{-1}$. After selecting events containing large-radius jets with high track multiplicity, the invariant mass distribution of the two highest-transverse-momentum jets is scanned to look for an excess above a data-driven estimate of the Standard Model multijet background. No significant excess of events is observed and the results are thus used to set 95\% confidence-level upper limits on the production cross-section times branching ratio of the $Z^\prime$ to dark quarks as a function of the $Z^\prime$ mass for various dark-quark scenarios.
}
 
\author{The ATLAS Collaboration}
 
\AtlasRefCode{HDBS-2018-45}
 
\PreprintIdNumber{CERN-EP-2023-217}

\AtlasJournalRef{JHEP 02 (2024) 128}
\AtlasDOI{10.1007/JHEP02(2024)128}

\hypersetup{pdftitle={ATLAS document},pdfauthor={The ATLAS Collaboration}}
 
\begin{document}
 
\maketitle

\section{Introduction}
\label{sec:intro}
 
In recent years,  exploring beyond the Standard Model (BSM) scenarios involving an additional non-Abelian hidden gauge group has been of increasing interest.
This sector, which would contain dark quarks and dark gluons, could provide dark matter candidates in the form of stable dark hadrons, leading to very peculiar signatures at colliders which have not yet been thoroughly explored.
For a review of such models and signatures see Ref.~\cite{Albouy:2022cin} and references therein.
 
At the Large Hadron Collider (LHC) dark quarks could be produced in the decay of a heavy mediator.
The dark quarks would then hadronise into dark hadrons, whose decay into Standard Model (SM) particles would produce objects which appear as jets with  properties that depend on the underlying theory parameters.
Limits were set for signatures which include jets aligned with missing transverse momentum (semi-visible jets) using the full Run~2 data samples by the ATLAS and CMS Collaborations~\cite{EXOT-2022-37,CMS-EXO-19-020}, and for jets with displaced tracks (emerging jets) using partial Run~2 data samples by the CMS Collaboration~\cite{CMS-EXO-18-001}.
 
\begin{figure}[h!]
\centering
\includegraphics[width=0.4\textwidth]{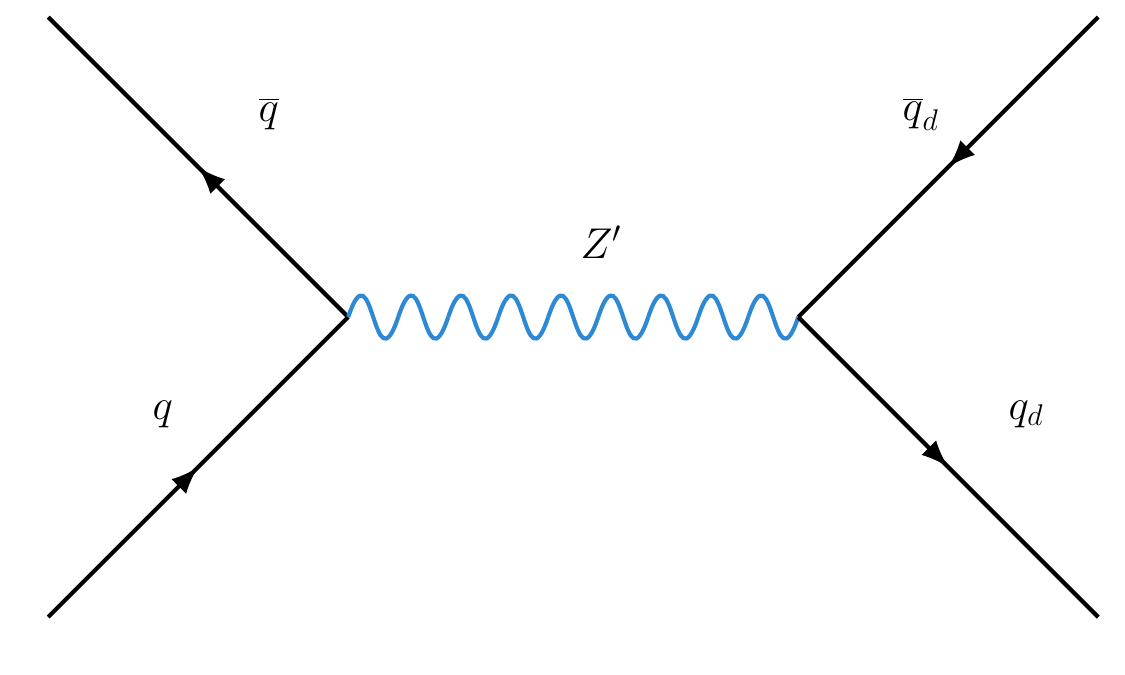}
\caption{Resonant dark quark pair production through a \Zprime mediator.}
\label{fig:feynman}
\end{figure}

In this paper, the resonant production of a pair of dark quarks, $q_d$, through a heavy \Zprime mediator is considered, as shown in Figure~\ref{fig:feynman}. In the benchmark models considered here, following Ref.~\cite{Park:2017rfb}, the dark hadrons are assumed to decay promptly into SM particles and the fraction of invisible components (stable dark hadrons) produced is assumed to be negligible, thus covering a complementary part of the parameter space relative to semi-visible jet and emerging jet searches, whose sensitivity to fully visible jets is reduced by requirements on the presence of missing transverse momentum. The dark jets considered here are typically wider than the SM QCD jets due in part to the double hadronisation process which would take place, first in the dark sector and then in the SM, after the dark mesons decay in part to high-momentum SM quarks.
While tracks associated to SM gluon and quark jets mostly come from charged pions and are thus directly linked to fragmentation, the charged particles inside the dark jets come from dark meson decays. As discussed in Ref.~\cite{Park:2017rfb}, the number of tracks associated to dark jets is thus influenced on the one hand by the multiplicity of dark mesons produced in the dark shower, which depends on the dark fragmentation and dark meson masses, and on the other hand by the decay channels of these dark mesons into SM particles. For the dark sector models considered here, this leads to a higher charged-particle multiplicity for dark jets than for SM jets.
This analysis hence looks for a resonant excess above a smooth background in the dijet invariant mass distribution, after preselecting events containing at least two high transverse momentum (\pt) large-radius jets with high track multiplicity, thus complementing the $\Zprime\rightarrow q\bar{q}$ dijet resonance search~\cite{EXOT-2019-03}. The invariant mass spectrum of the dominant SM QCD dijet background is determined using a data-driven method.
 
\section{ATLAS detector}
\label{sec:detector}
\newcommand{\AtlasCoordFootnote}{
ATLAS uses a right-handed coordinate system with its origin at the nominal interaction point (IP)
in the centre of the detector and the \(z\)-axis along the beam pipe.
The \(x\)-axis points from the IP to the centre of the LHC ring,
and the \(y\)-axis points upwards.
Cylindrical coordinates \((r,\phi)\) are used in the transverse plane,
\(\phi\) being the azimuthal angle around the \(z\)-axis.
The pseudorapidity is defined in terms of the polar angle \(\theta\) as \(\eta = -\ln \tan(\theta/2)\).
Angular distance is measured in units of \(\Delta R \equiv \sqrt{(\Delta\eta)^{2} + (\Delta\phi)^{2}}\).}
 
The ATLAS experiment~\cite{PERF-2007-01} at the LHC is a multipurpose particle detector
with a forward--backward symmetric cylindrical geometry and a nearly \(4\pi\) coverage in
solid angle.\footnote{\AtlasCoordFootnote}
It consists of an inner tracking detector surrounded by a thin superconducting solenoid
providing a \qty{2}{\tesla} axial magnetic field, electromagnetic and hadron calorimeters, and a muon spectrometer.
The inner tracking detector covers the pseudorapidity range \(|\eta| < 2.5\).
It consists of silicon pixel, silicon microstrip, and transition radiation tracking detectors.
Lead/liquid-argon (LAr) sampling calorimeters provide electromagnetic (EM) energy measurements
with high granularity.
A steel/scintillator-tile hadron calorimeter covers the central pseudorapidity range (\(|\eta| < 1.7\)).
The endcap and forward regions are instrumented with LAr calorimeters
for both the electromagnetic and hadronic energy measurements up to \(|\eta| = 4.9\).
The muon spectrometer surrounds the calorimeters and is based on
three large superconducting air-core toroidal magnets with eight coils each.
The field integral of the toroids ranges between \num{2.0} and \qty{6.0}{\tesla\metre}
across most of the detector.
The muon spectrometer includes a system of precision tracking chambers and fast detectors for triggering.
 
A two-level trigger system is used to select events.
The first-level trigger is implemented in hardware and uses a subset of the detector information
to accept events at a rate below \qty{100}{\kHz}.
This is followed by a software-based trigger that
reduces the accepted event rate to \qty{1}{\kHz} on average.
 
An extensive software suite~\cite{ATL-SOFT-PUB-2021-001} is used in simulation, in the reconstruction
and analysis of real and simulated data, in detector operations, and in the trigger and data acquisition
systems of the experiment.
 
\section{Data and simulated events}
\label{sec:dataMC}
 
The data sample used in this analysis was collected using unprescaled single-jet triggers, where jets were reconstructed exploiting the \akt clustering algorithm~\cite{Cacciari:2008gp, Fastjet} with a radius $R$ = 1.0 and a \pT threshold of 460 ~\GeV, with varying jet trimming~\cite{Krohn:2009th} requirements depending on the year of data taking. These triggers are fully efficient after applying the offline object requirements described in the next section. The integrated luminosity of the data sample is 139~\ifb after applying the stable beam conditions, detector and data quality requirements~\cite{DAPR-2018-01}.
 
The background from multijet processes is simulated with the Pythia 8.186~\cite{Sjostrand:2007gs} event generator. The A14 set of tuned parameters~\cite{ATL-PHYS-PUB-2014-021} is used together with the NNPDF2.3LO PDF (parton distribution function) set~\cite{Ball:2012cx}. The contributions of other processes is found to be negligible after selections and not considered further in the analysis.
 
Signal samples of $pp\rightarrow\Zprime\rightarrow q_d\bar{q}_d$, considering an SU(3) symmetry in the dark sector, are simulated using the Hidden Valley module of Pythia 8.235~\cite{Sjostrand:2014zea} with \Zprime masses varying from 1.5 to 3.5 ~\TeV, using the same set of tuned parameters and PDF set as the multijet samples.
Between fully simulated points, which are spaced by 250 or 500~\GeV, the signal acceptance and shape are interpolated.
The production cross-section is a free parameter as it depends on the coupling of the \Zprime to SM quarks, $g_q$. This parameter is set indirectly by choosing the \Zprime width and its relative branching ratio to SM quarks.
For the cross-section values explored in this paper, the  scaling $g_q$ does not affect the dijet mass distribution because the true \Zprime width remains much smaller than the reconstructed width that is dominated by resolution and reconstruction effects.
To probe the dark QCD sector, four combinations of parameters are chosen to cover different possibilities for the number of dark quark flavours, $n_f$, the confinement scale $\Lambda_d$, the mass spectra and the prompt decay modes of the dark pions, $\pi_{d}$.
The values of these parameters are chosen following the benchmark models proposed in Ref.~\cite{Park:2017rfb}, and are summarised in Table~\ref{tab:signal}.
While the decay of the dark rho mesons is fixed to $\rho_d \to \pi_d \pi_d$, the dark pions are either assumed to decay directly to SM quarks (models A and B) or to SM particles through dark photons, $\gamma'$ (models C and D).
In the latter case, the dark photons decay according to the branching ratios computed in Ref.~\cite{Buschmann:2015awa}.
In model C, the dark photons decay into any pair of charged fermions excluding third-generation quarks (due to the choice of dark photon mass), while in model D the dark photons decay predominantly into charged pions, and occasionally to electron and muon pairs.
 
\begin{table*}[htp!]
\begin{center}
\caption{Benchmark models taken from the reference in the text (Ref.~\cite{Park:2017rfb}). In the headings, $n_f$ corresponds to the number of low-mass dark quarks flavors, $\Lambda_d$ is the dark confinement scale and the masses $\tilde{m}_{q'}$, $m_{\pi_d}$, $m_{\rho_d}$ correspond to the constituent quark mass---the masses of the dark pion and dark rho mesons respectively.}
\begin{tabular}{|c|c|c|c|c|c|p{2.7cm}|p{3cm}|}
\hline \multirow{ 2}{*}{Model} & \multirow{ 2}{*}{$n_f$} & $\Lambda_d$ &  $\tilde{m}_{q'}$ & $m_{\pi_d}$ & $m_{\rho_d}$ &
\multirow{ 2}{*}{$\pi_d$ decay mode}\\
& & (\GeV) &  (\GeV) & (\GeV) & (\GeV) &  \\
\hline$A$ &2&15&  20  &10&50&$\pi_d \to c \bar{c}$\\
\hline$B$ &6&2&  2 &2&4.67&$\pi_d \to s \bar{s}$ \\
\hline \multirow{ 2}{*}{$C$} &\multirow{ 2}{*}{2}&\multirow{ 2}{*}{15}& \multirow{ 2}{*}{20} &\multirow{ 2}{*}{10}&\multirow{ 2}{*}{50}&$\pi_d \to \gamma' \gamma' $ {with $m_{\gamma'} = 4.0~\GeV$}   \\
\hline \multirow{ 2}{*}{$D$} &\multirow{ 2}{*}{6}&\multirow{ 2}{*}{2}&\multirow{ 2}{*}{2} &\multirow{ 2}{*}{2}&\multirow{ 2}{*}{4.67}&$\pi_d \to \gamma' \gamma' $ {with $m_{\gamma'} = 0.7~\GeV$} \\
\hline\end{tabular}
\label{tab:signal}
\end{center}
\end{table*}
 
For all simulated samples, the EvtGen 1.2.0 program~\cite{Lange:2001uf} is used for the properties of $b$- and $c$-hadron decays. The generation of the simulated event samples also includes the effect of multiple $pp$ interactions per bunch crossing, and the effect on the detector response due to interactions from bunch crossings before or after the one containing the hard interaction. Multiple overlaid $pp$ collisions are simulated with the soft QCD processes of Pythia 8.186~\cite{Sjostrand:2014zea} using the A2 set of tuned parameters ~\cite{ATL-PHYS-PUB-2012-003} and the MSTW2008LO PDF set~\cite{Martin:2009iq}. All the Monte Carlo  events are processed through a full simulation of the ATLAS detector geometry and response~\cite{SOFT-2010-01} based on  \GEANT~\cite{Agostinelli:2002hh}.
 
\section{Object and event selections}
\label{sec:object}
 
Events are required to have at least one $pp$ interaction vertex with at least two tracks with \pT greater than 500~\MeV. The primary vertex is defined as the one with the largest $\Sigma\pT^2$ of its associated tracks. Events are rejected if they contain jets compatible with noise bursts, beam-induced background or cosmic rays as defined in Ref.~\cite{ATLAS-CONF-2015-029}.
 
Large-$R$ jets are built from three-dimensional topological clusters of energy deposits in the calorimeter, calibrated to the hadronic energy scale with the local cluster weighting (LCW)~\cite{PERF-2014-07} procedure using the \akt algorithm~\cite{Cacciari:2008gp,Fastjet} with a radius parameter $R$ = 1.0. To reduce contributions from pile-up, initial-state radiation and multiple parton interactions that would contaminate the measurement of the hard scatter process, the jets are trimmed~\cite{Krohn:2009th}. The trimming procedure reclusters jet constituents into subjets using the \kt algorithm~\cite{Catani:1991hj,Ellis:1993tq,Catani:1993hr} with $R$ = 0.2 and discards the resulting subjets if their  \pT is less than 5\% of the \pT of the parent jet~\cite{ATLAS-CONF-2015-035},  before recomputing the four-momenta of the large-$R$ jets using the remaining subjets.
 
This analysis selects events containing at least two large-$R$ jets satisfying the following requirements: $500 < \pT < 3000$~\GeV~for the leading and $400 < \pT < 3000$~\GeV~for the subleading jet, $|\eta| < 2.0$ and $50 < m_\mathrm{J} < 600$~\GeV, where $m_\mathrm{J}$ is the invariant mass of the jet. The upper bounds on \pT and $m_\mathrm{J}$ have a small impact on the signal acceptance while ensuring that jet energy and mass scale calibrations and their uncertainties remain valid~\cite{JETM-2018-02,ATLAS-CONF-2020-022}. Finally, along with the \pT$>500$~\GeV selection on the leading jet, \mjj is required to be greater than 1.3~\TeV~to ensure full trigger efficiency.
 
The distributions of the dijet invariant mass of the multijet background and benchmark models with $m_\Zprime = 2.5$~\TeV are shown in Figure~\ref{fig:preselmjj} after these preselections. The bin width in \mjj is determined by the experimental dijet mass resolution (evaluated on background simulation) of approximately 100~\GeV~for the background in this mass range.
The signal distributions are relatively broad and shifted to lower values relative to the true \Zprime mass. This is mainly due to the removal of some of the jet constituents by the trimming algorithm, which also reduces the background under the bulk of the signal peak~\cite{Krohn:2009th}.
 
\begin{figure}[h!]
\centering
\includegraphics[width=0.49\textwidth]{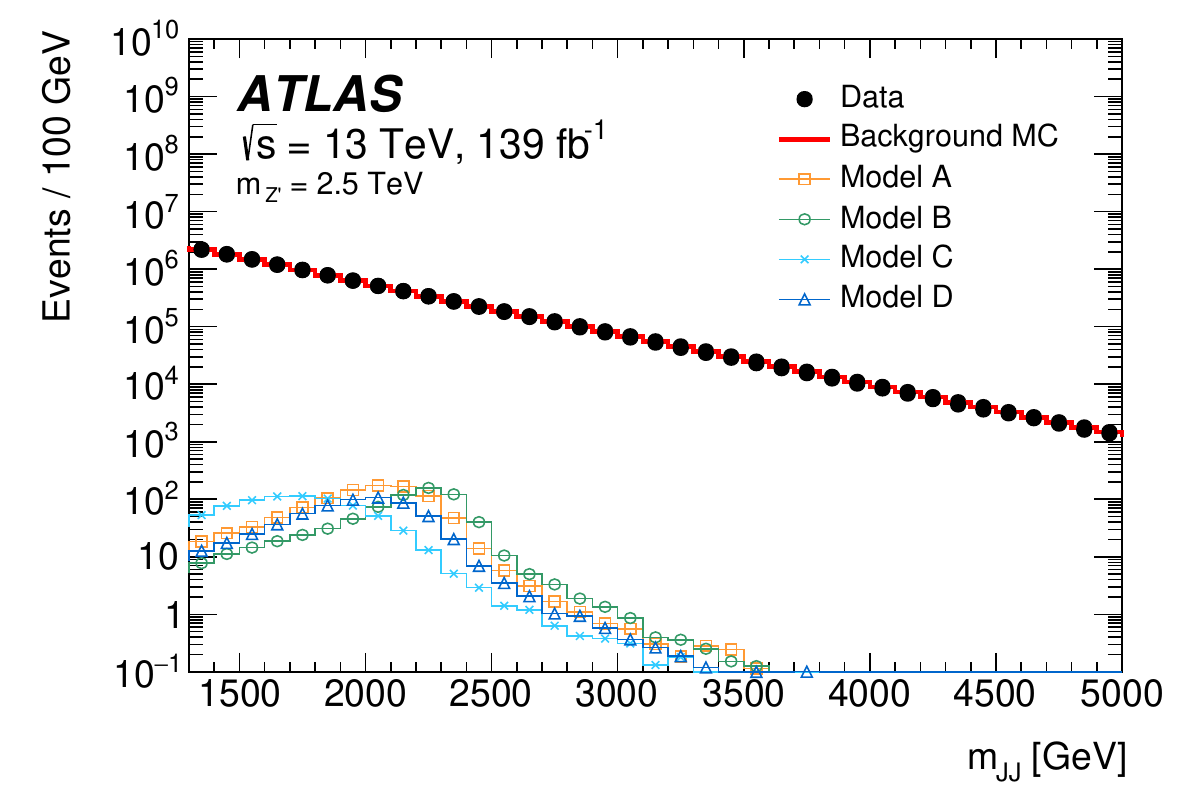}
\caption{Distribution of the dijet invariant mass, \mjj for the data, the simulated multijet background and of some representative signals (models A, B, C and D with $m_{\Zprime}=2.5$~\TeV), shown after applying the preselections described in the text.  The simulated background is normalised to the data and the signals are normalised to a production cross-section of 10 fb.}
\label{fig:preselmjj}
\end{figure}
 
The number of tracks matched to the untrimmed jets via ghost association\footnote{Ghost association of tracks is a method that associates tracks to jets by giving them negligible momentum and clustering them within the jets using the jet reconstruction algorithm.}~\cite{Cacciari:2008gn}, \ntrk, is helpful in separating the signal from the background. These tracks are required to have $\pT > 500$ ~\MeV, fulfil hit requirements corresponding to the \emph{Loose} working point~\cite{ATL-PHYS-PUB-2015-051} and to be matched to the primary vertex.\footnote{This matching requires that the difference between the track longitudinal impact parameter and the primary vertex when expressed at the beam line ($|z_0 \sin{\theta}|$) is less than 3 mm (and likewise with respect to any other vertex in the event) and that the track transverse impact parameter $d_0$ is less than 2 mm.}
The distributions of \ntrk for the multijet background and representative signals are shown in Figure~\ref{fig:preselntrk} after the preselections for the two leading jets are applied. In the high end of the spectrum where the selection is made, the data agree with the background prediction within the modelling uncertainties that are described in Section~\ref{sec:syst}. It was checked that the small fluctuation found at very low values of \ntrk has a negligible impact on the analysis.
 
\begin{figure}[htp!]
\begin{center}
\subfigure[]{\includegraphics[width=0.45\linewidth]{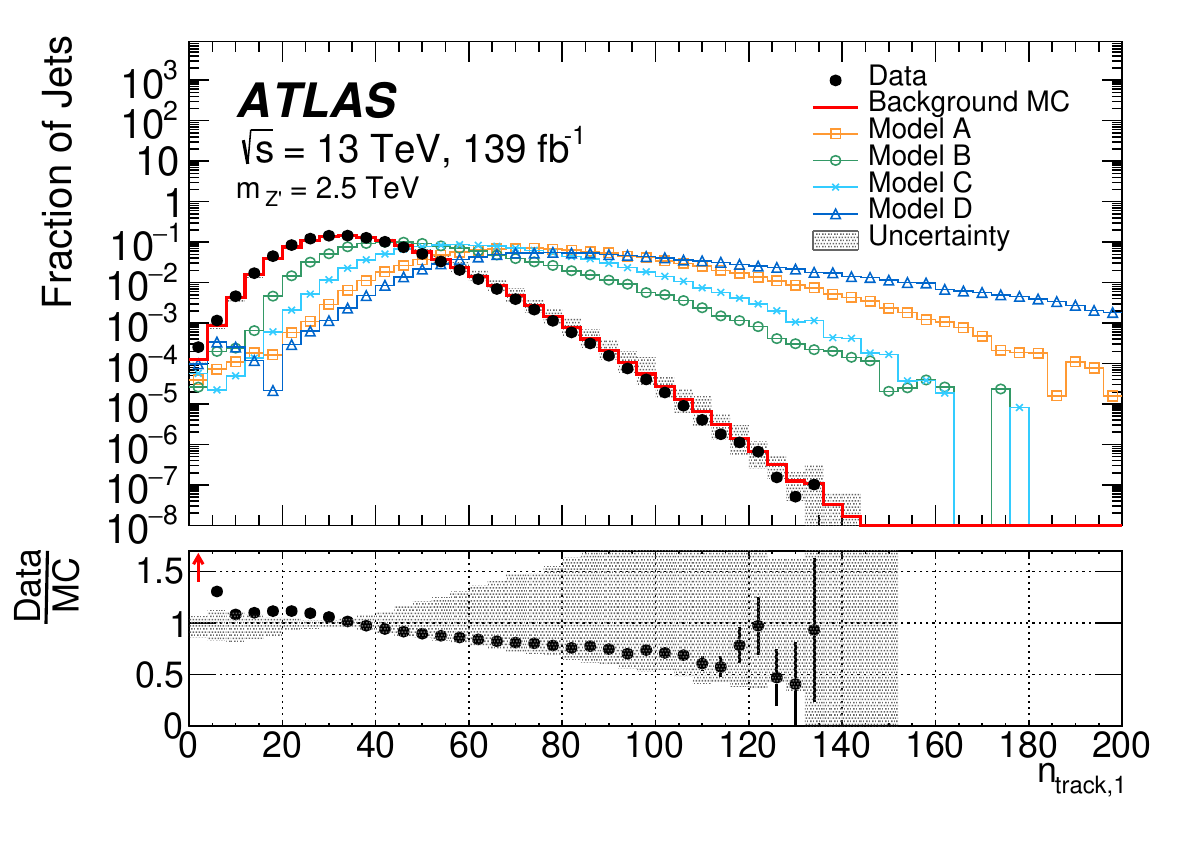}\label{fig:leadjet_logy}}
\subfigure[]{\includegraphics[width=0.45\linewidth]{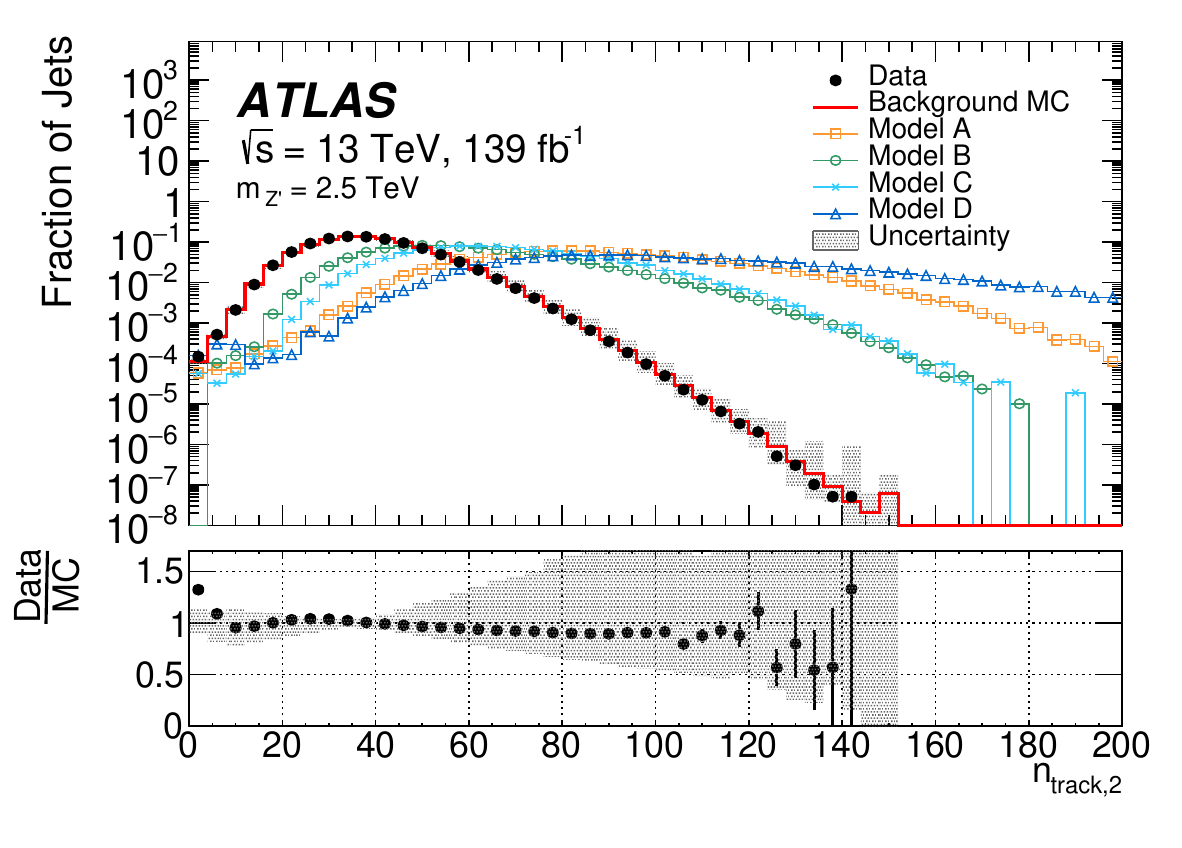}\label{fig:subjet_logy}} \\
\caption{Distributions of the number of tracks associated with the two leading jets, \ntrkone (a) and \ntrktwo (b), for the data, the simulated multijet background and of some representative signals (models A, B, C and D with $m_{\Zprime}=2.5$ ~\TeV), shown after applying the preselections described in the text.  All distributions are normalised to unity. The uncertainty band around the background prediction corresponds to the modelling uncertainty described in Section~\ref{sec:syst}. A red arrow indicates an out-of-range data point for a given bin.}
\label{fig:preselntrk}
\end{center}
\end{figure}

To enhance the signal-to-background ratio, it is beneficial to require that the jets have a minimal value of \ntrk . However, such a requirement would sculpt the \mjj background distribution since the number of tracks associated with a jet is correlated to its \pt . To avoid this undesirable effect, which would impact the identification of a clear resonance over the background spectrum, a new variable, \ntrke, is defined as follows:
\begin{itemize}
\item{First, a target efficiency, $\epsilon$, for a background jet to pass the requirement on \ntrk is defined. In this analysis, a value of $\epsilon=1\%$ is chosen to reduce the background significantly while maintaining good acceptance across the various signals probed.}
\item{Second, for each bin in \mjj and for the leading jet of the simulated multijet background, the minimal value \PJone
for which \ntrkone > \PJone leads to the target efficiency, $\epsilon$, is derived (where the subscript ``1'' refers to the leading jet).}
\item{Third, the procedure is repeated with the subleading jet to determine \PJtwo, after applying \ntrkone $>$ \PJone on the leading jet. This is done to account for possible correlation effects between the two leading jets. Applying an \ntrk requirement on the leading jet selects preferentially events which also contain higher-\ntrk subleading jets, raising the value of \PJtwo needed to reach the target efficiency. }
\item{Fourth, the  \PJi values obtained for the two leading jets ($i=1,2$) are fitted as a function of \mjj  to have a smooth distribution. The values obtained for \PJone and \PJtwo are shown in Figure~\ref{fig:percentiles}.}
\item{Finally, the variable $\ntrkei$ is defined for the two leading jets as $\ntrkei = \ntrki - \PJi$. Requiring $\ntrkeone>0$ removes $1-\epsilon$ of the multijet background; requiring $\ntrketwo>0$ further removes $1-\epsilon$ of the remaining background.}
\end{itemize}
 
\begin{figure}[h!]
\centering
\includegraphics[width=0.49\textwidth]{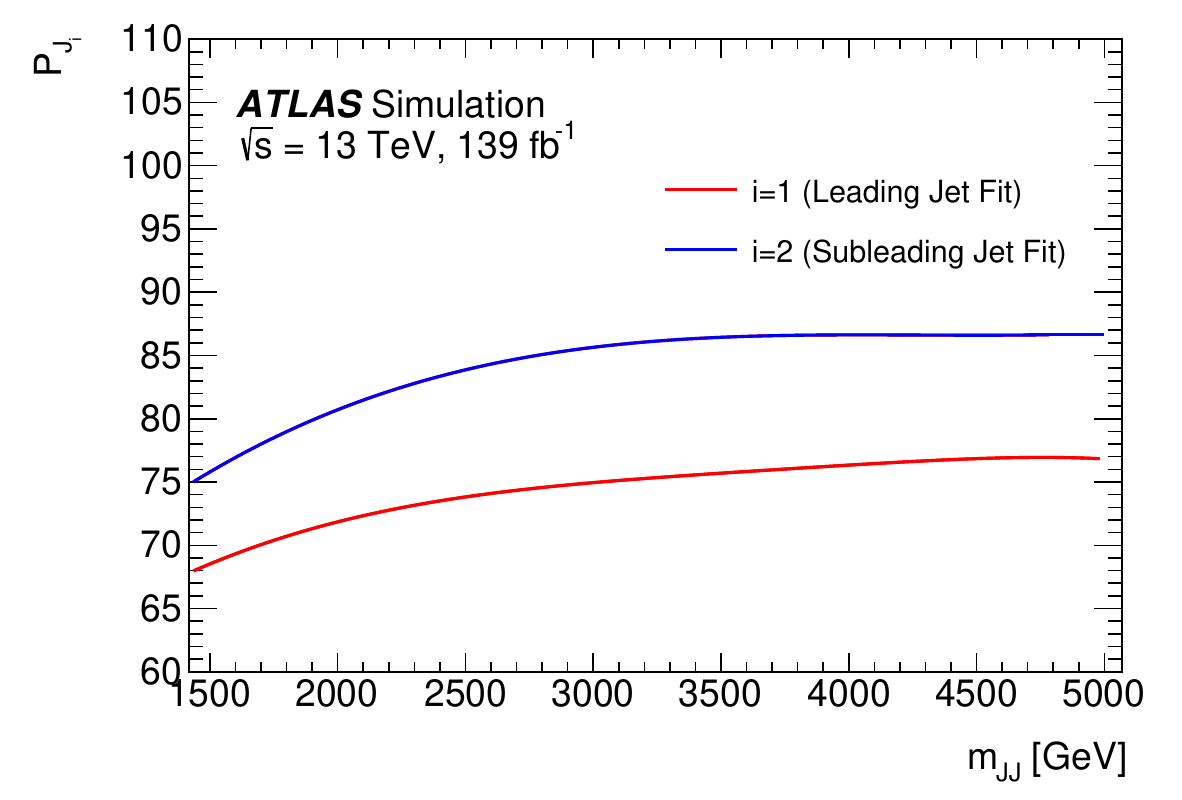}
\caption{Distributions of \PJone (bottom curve) and \PJtwo (top curve) as a function of \mjj for the simulated multijet background.
These variables are used to define a selection on \ntrk which evolves with \mjj in order to achieve a fixed background selection efficiency as a function of \mjj.
As mentioned in the text, the value of \PJtwo is determined after applying
\ntrkone $<$ \PJone on the leading jet,
thus preferentially selecting events which also contain higher-\ntrk subleading jets.
}
\label{fig:percentiles}
\end{figure}
 
The signal region (SR) is defined by requiring $\ntrkeone > 0$ and $\ntrketwo > 0$ after the preselections are applied, which reduces the multijet background by approximately 99.99\%.
This selection also reduces the already-negligible backgrounds from top and electroweak processes ($<$ 1\% compared to the multijet background).
For $m_\Zprime=1.75$~\TeV the signal acceptance times efficiency after all selections lies between 0.3\% and 9\%, while at $m_\Zprime=2.5$ (3.5)~\TeV it is between 1\% (2.5\%)  and 28\% (37\%) depending on the signal models.
 
\section{Background estimation}
\label{sec:bgestimate}
 
The requirement on $\ntrke$ can also be reversed by requiring $\ntrke < 0$ for the two leading jets, in order to construct a control region (CR) which is largely enriched in multijet background while maintaining the same \mjj shape, as per definition $\ntrke$ is decorrelated from \mjj . This allows a data-driven estimate of the background in the signal region. The background contribution to the SR is determined by taking the \mjj distribution directly from the CR, while the overall background normalisation is left as a free parameter to be extracted from a likelihood fit in the SR. This CR method is adopted for the background estimate because of the relatively large reconstructed width of the signal models considered. Signal injection tests, in which signal models are added to the data-driven expected background distribution in the SR, show that the extracted signal yields obtained when using the CR method are consistent with the injected values within uncertainties. The background shape template from the CR can create a bias in the likelihood fit, a ``spurious signal,'' $S_\mathrm{spur}$, which corresponds to the extracted number of signal events when a signal-plus-background fit is conducted on a background-only sample. A spurious signal test is thus conducted and is shown to satisfy the criterion motivated in Ref.~\cite{ATL-PHYS-PUB-2020-028} for all models considered. The requirement  that $S_\mathrm{spur}/\sigma_\mathrm{fit}<0.5$ is used  where $\sigma_\mathrm{fit}$ is the uncertainty in the fitted number of signal events. In practice it is found to be less than 0.35 for all the models and masses considered.
 
A validation region (VR), between the CR and the SR, is also built by requiring $\ntrkeone > 0$ and $\ntrketwo < 0$. This region allows the verification that the decorrelation of \ntrke and \mjj, derived from the simulated multijet processes, operates well in data: in the VR, the scaled CR template agrees with the data with a chi-squared per degree of freedom of 1.2. This good agreement is shown in Figure~\ref{fig:CRvsVR} in which the \mjj data distribution in the CR is compared to that in the VR after normalising the total number of events in each region to the one expected in the SR. The general \mjj shape agrees well between the CR and the VR, which confirms that the decorrelation method works within the statistical uncertainty expected in the SR.
Both the CR and the VR have significantly more background events than the SR: the expected signal-to-background ratio in these regions is thus found to be negligible with respect to that in the SR, for all the models considered. Therefore, any bias from the presence of signal events in these regions will be negligible, in either the construction of the background model or in the above cross-checks.
 
\begin{figure}[h!]
\centering
\includegraphics[width=0.7\textwidth]{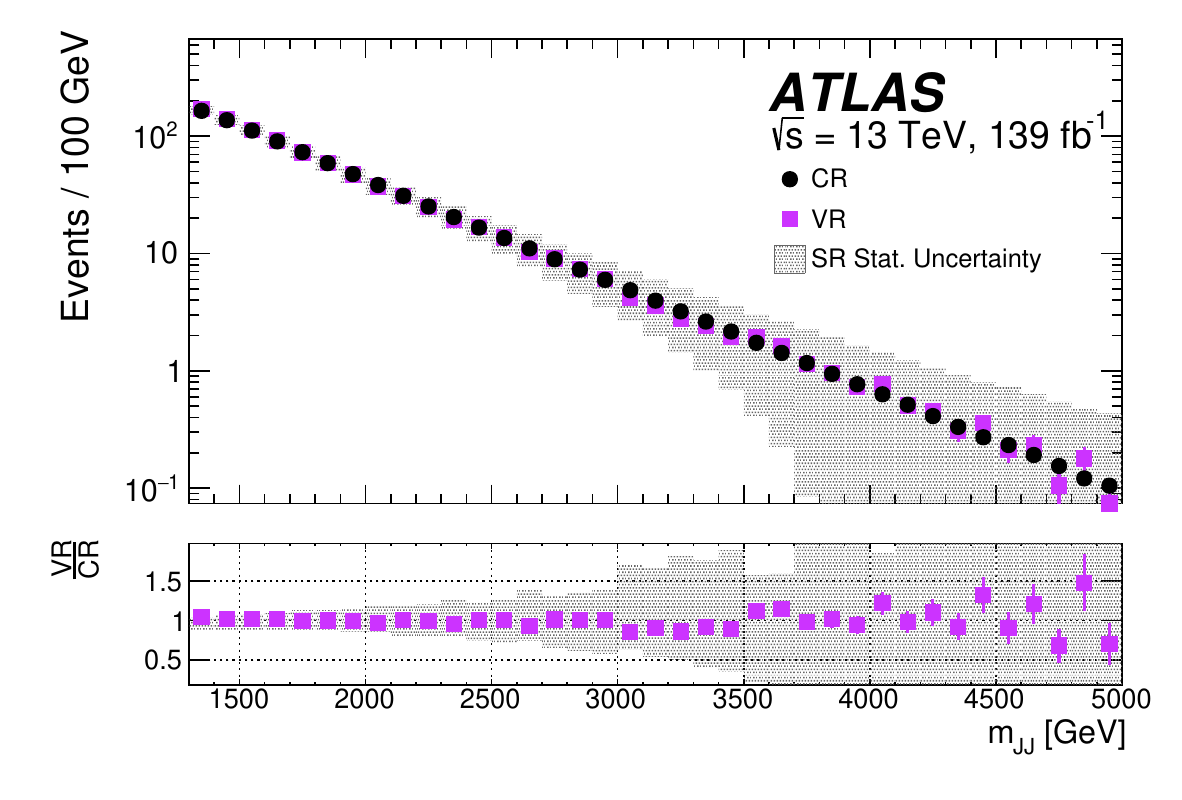}
\caption{Distribution of \mjj in the CR (black dots) compared with that in the VR (purple dots), with their respective statistical uncertainties, after normalising the total number of events in each region to the one expected in the SR. The uncertainties in the data points correspond to the uncertainties on the number of events in the CR and the VR, and the shaded grey band shows the statistical uncertainty expected in the SR.}
\label{fig:CRvsVR}
\end{figure}
 
The statistical analysis of the data follows a frequentist approach based on a profile-likelihood ratio test statistic~\cite{Cowan:2010js}. The parameter of interest---the signal strength $\mu$---is extracted from a binned maximum-likelihood fit of a signal-plus-background model to the \mjj distribution in the SR. The signal strength then acts as a scale factor on the predicted number of events for each model assumption and is allowed to take negative values in the fit. The bin width is fixed at 100~\GeV, as motivated by fitting the experimental dijet mass resolution of the background simulation,
while the \mjj range probed extends from 1.3 to 5~\TeV. The likelihood is defined from the Poisson probability to observe $N$ data events for a given expected signal $n_{s}$ and background $\nu n_{b}$, where $\nu$ is the background normalisation scale factor that is extracted from the fit, as:
\begin{equation}
\mathcal{L}=\prod_{i}^{bins}\mathrm{Poiss}(N_{i}|\mu n_{s,i}(\vec{\theta})+\nu n_{b,i}(\vec{\theta}))\prod_{j \in \vec{\theta}}G_{j}(\theta_j), \nonumber
\end{equation}
where the nuisance parameters, $\vec{\theta}$, representing the systematic uncertainties $j$, are included with Gaussian constraint terms $G$.  The systematic uncertainties are detailed in Section~\ref{sec:syst}. The expected background shape, $n_{b,i}$ is taken from the CR.

\section{Systematic uncertainties}
\label{sec:syst}
 
Systematic uncertainties in the background estimate and in the expected signal yields and shapes arise from detector effects and Monte Carlo (MC) modelling; they are treated as nuisance parameters in the statistical analysis. Given that the shape of the background is taken from the CR and that its normalisation is a free parameter determined by a fit in the SR, for all  the uncertainties listed below, only their effects on the shape agreement between the SR and the CR are considered as uncertainties in the background.
 
The experimental uncertainties include uncertainties on the jets, tracks and luminosity.
The uncertainties associated with the large-$R$ jets arise from the jet energy and mass scales, mass response and energy resolution~\cite{JETM-2018-02,JETM-2018-05}.
For the model B, C and D simulated signal samples, an additional jet energy scale uncertainty of 5\% is added to cover for a small deviation from unity in the jet energy response with respect to the true energy scale (non-closure) observed after calibration following the SR selections. For model A and for the SM dijet background, closure was obtained within the nominal uncertainties. Uncertainties in the track reconstruction performance are related to their reconstruction efficiency and to the rate at which fake tracks are reconstructed~\cite{ATL-PHYS-PUB-2015-051}. These also include additional uncertainties related to tracking in dense environments, given the large number of tracks requested in the core of high-\pT jets in this analysis~\cite{PERF-2015-08}.
The limited tracker coverage in combination with the use of large-$R$ jets does not affect the results, since its effect is found to be small and well modelled within the CR and the VR\@.
The uncertainty in the combined 2015--2018 integrated luminosity, only affecting the signal samples, is 1.7\%~\cite{ATLAS-CONF-2019-021}, obtained using the LUCID-2 detector~\cite{Avoni:2633501} for the
primary luminosity measurements.
 
The MC generator uncertainties are applied to signal and include uncertainties in the PDFs~\cite{Butterworth:2015oua}, renormalisation and factorisation scales, and strong coupling constant, as well as uncertainties related to the parton shower set of tuned parameters in initial- or final-state radiation, such as variations of the renormalisation scale for QCD emission ($\mu_{R,FSR}$) or the inclusion of non-singular terms~\cite{Mrenna:2016sih}.
 
Following the prescription described in Ref.~\cite{ATL-PHYS-PUB-2020-028}, an uncertainty related to the spurious signal test is added. It is estimated by performing a signal-plus-background fit on the VR data which is scaled down to the expected number of events in the SR.
 
The statistical uncertainty dominates over the systematic uncertainties over the full mass range probed.
The impact of the largest relative systematic uncertainties on $\mu$ are shown in Table~\ref{tab:systeffect} for the four models and an example \Zprime mass of 2.5~\TeV.
They are determined by re-running the fit while  fixing, in turn, each of these nuisance parameters to their post-fit value plus or minus one standard deviation. Models B and C are most affected by uncertainties which can shift signal events in and out of acceptance and, due to a lower number of tracks and a more displaced $m_\mathrm{JJ}$ peak, these models have the lowest acceptance.
 
\begin{table*}[htp!]
\begin{center}
\caption{Impact of the largest relative systematic uncertainties, $\Delta\mu/\mu$ (in percentage), for the various models and for $m_\Zprime=2.5$~\TeV.}
\begin{tabular}{l|c c c c }
\hline & \multicolumn{4}{c}{Model}\\
\hline Uncertainty & A & B &  C & D  \\\hline
$\mu_{R,FSR}$ & 7.3 & 19.0 & 34.1 & 9.9 \\
Jet calibration non-closure & -- & 25.6 & 27.3 & 13.8 \\
Spurious signal & 10.7 & 14.7 & 3.7 & 10.3 \\
PDF & 4.9 & 5.5 & 4.8 & 4.8 \\
\hline\end{tabular}
\label{tab:systeffect}
\end{center}
\end{table*}

\section{Results}
\label{sec:results}
 
Figure~\ref{fig:mjjSR} shows the observed \mjj distribution in the SR\@. The highest mass event in data is observed at 4.8~\TeV.  A BumpHunter test~\cite{Choudalakis:2011qn} is performed on the distribution to test its compatibility with the background-only hypothesis: the vertical lines on the Figure indicate the most discrepant interval identified which has a global $\textrm{p}_0$-value of $0.63$ in the interval 1500~\GeV $< \mjj < 1700$~\GeV. This $\textrm{p}_0$-value and the bin-by-bin significance of the difference between the data and the fit, shown in the lower panel of the plot, consider only statistical uncertainties.
 
\begin{figure}[htp!]
\centering
\includegraphics[width=0.7\textwidth]{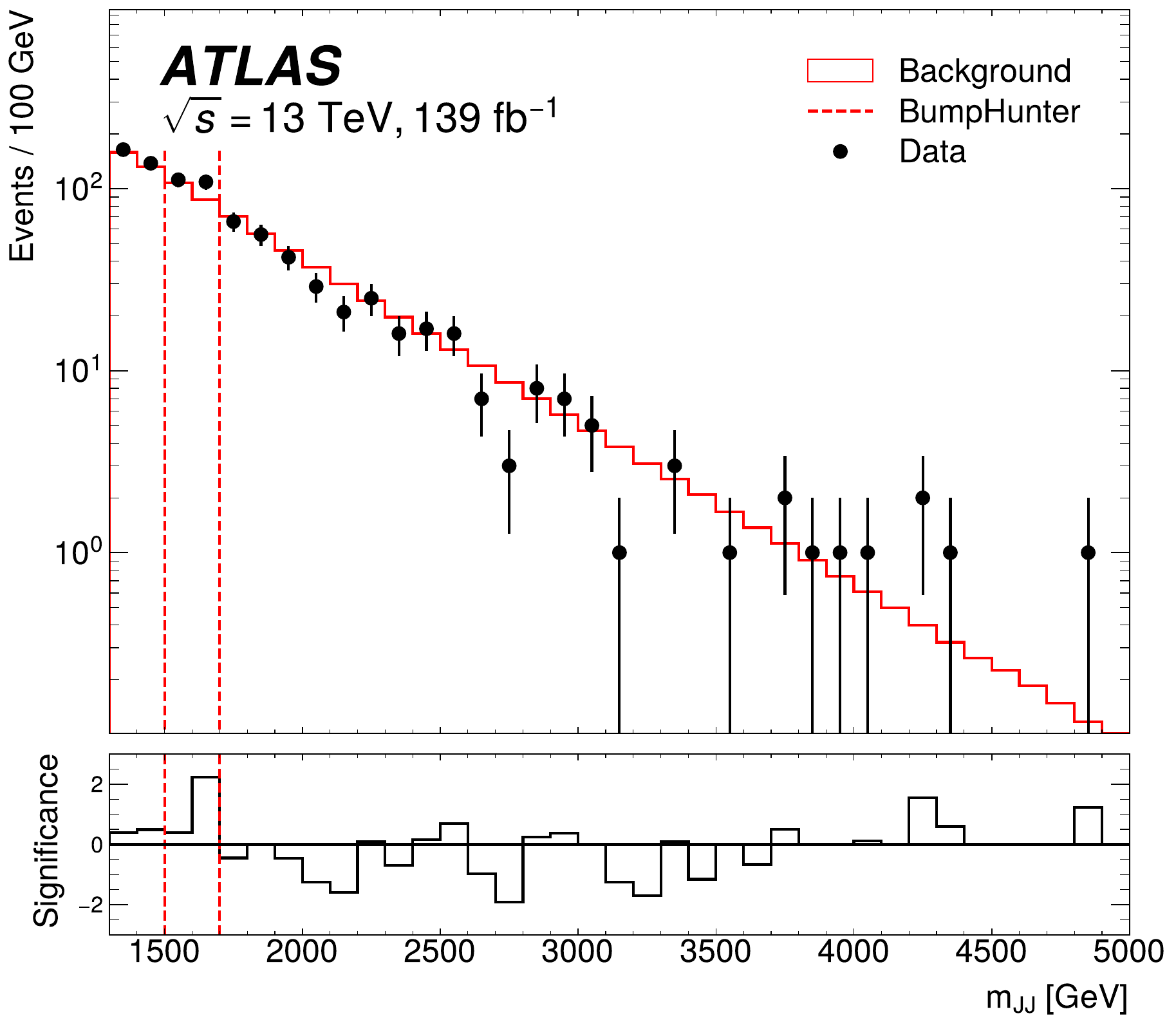}
\caption{Distribution of \mjj in the SR (black points) compared to the background-only expectation, where the shape is derived from the CR and the normalisation is determined from a fit in the SR. The most discrepant interval found by the BumpHunter algorithm, indicated by the vertical lines, is between 1500~\GeV < \mjj < 1700~\GeV, with a $\textrm{p}_0$-value of $0.63$. This $\textrm{p}_0$-value and the bin-by-bin significance shown in the lower panel are computed considering statistical uncertainties only.}
\label{fig:mjjSR}
\end{figure}
 
Given no significant deviation from the expected background is observed in Figure~\ref{fig:mjjSR}, limits can be set on the various models considered as shown in Figure~\ref{fig:limits}: upper limits on the signal production cross-section times branching ratio of the \Zprime to dark quarks are extracted at 95\% confidence level (CL) using the CL$_\text{s}$ method~\cite{Read:2002hq}, with the asymptotic formulae\footnote{The expected limits using the asymptotic formula and using 10000 pseudo-experiments are similar for most mass points and models, with differences below 10-20 \%, except for model B above a \Zprime mass of 3.4~\TeV, for which the difference goes up to 35\%. This thus only affects the two highest-mass points of model B as this model displays the highest reconstructed \mjj for a given generated $m_{\Zprime}$.} for the profile likelihood ratio test statistic distribution~\cite{Cowan:2010js}. The \Zprime-mass lower bound in the limit plots is chosen such that the mean of the signal \mjj distribution is above the lowest \mjj value probed of 1.3~\TeV. Given the \mjj shape of the data in the SR, the signal-plus-background fit is able to accommodate some large-width, low-$m_{\Zprime}$ signal by lowering the background normalisation at low-\mjj values, while still being compatible with the data in the higher-\mjj region. The converse is true for higher-$m_{\Zprime}$ values, resulting in the observed limit shape.
 
In Figure~\ref{fig:limits}, the limits are also compared to theory predictions for given values of $g_q$ and $g_{q_d}$; these are obtained at leading-order in QCD with \MGNLO~\cite{Alwall:2014hca}, using a model based upon that used in the $\Zprime\rightarrow q\bar{q}$ ATLAS dijet resonance search~\cite{EXOT-2019-03}, in which the coupling to dark matter particles is replaced by a coupling to dark quarks. In that model, a universal coupling of the $Z'$ to each SM quark $g_q$ is assumed and the dark quarks are all represented by one species with an effective coupling $g_{q_d}$ to the $Z'$. For $g_q=0.05$ and $g_{q_d}=0.2$, a value of $g_q$ which evades the dijet resonance search constraints, the observed (expected) limit on the \Zprime mass  is at 3.0 (2.9)~\TeV~for model A and reaches up to 2.9 (2.8)~\TeV~for model D (for which the region between 2.1 and 2.2~\TeV is not excluded), while for models B and C, due to a lower acceptance, these couplings do not lead to any exclusion. Pushing the couplings up to $g_q=0.15$ and $g_{q_d}=0.5$, a regime in which the \Zprime width is still below the experimental resolution of about 100~\GeV, \Zprime masses up to 2.3~\TeV~and between 2.8 and 3.5~\TeV~can be excluded for model C, a sensitivity which is close to the one obtained by the dijet resonance analysis which sets a limit for $g_q=0.15$ at around 3.5~\TeV.
 
\begin{figure}[htp!]
\begin{center}
\subfigure[]{\includegraphics[width=0.45\linewidth]{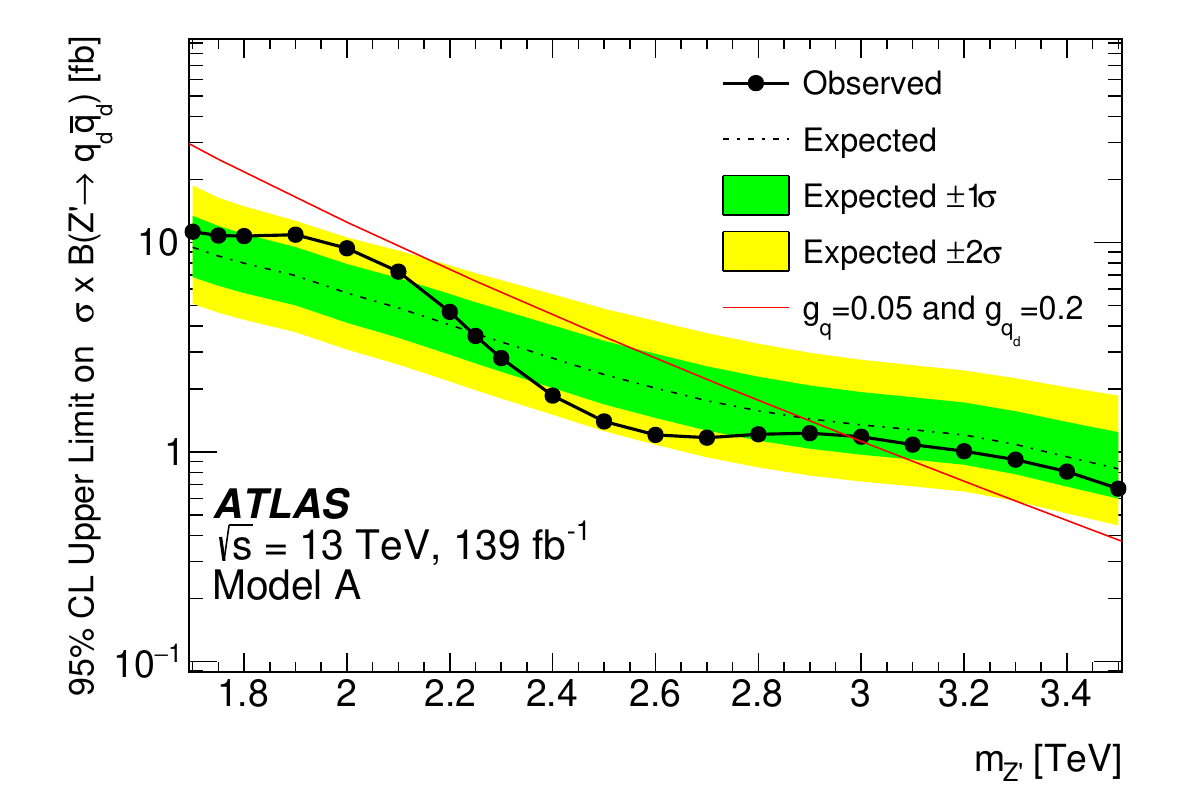}\label{fig:limits_modelA}}
\subfigure[]{\includegraphics[width=0.45\linewidth]{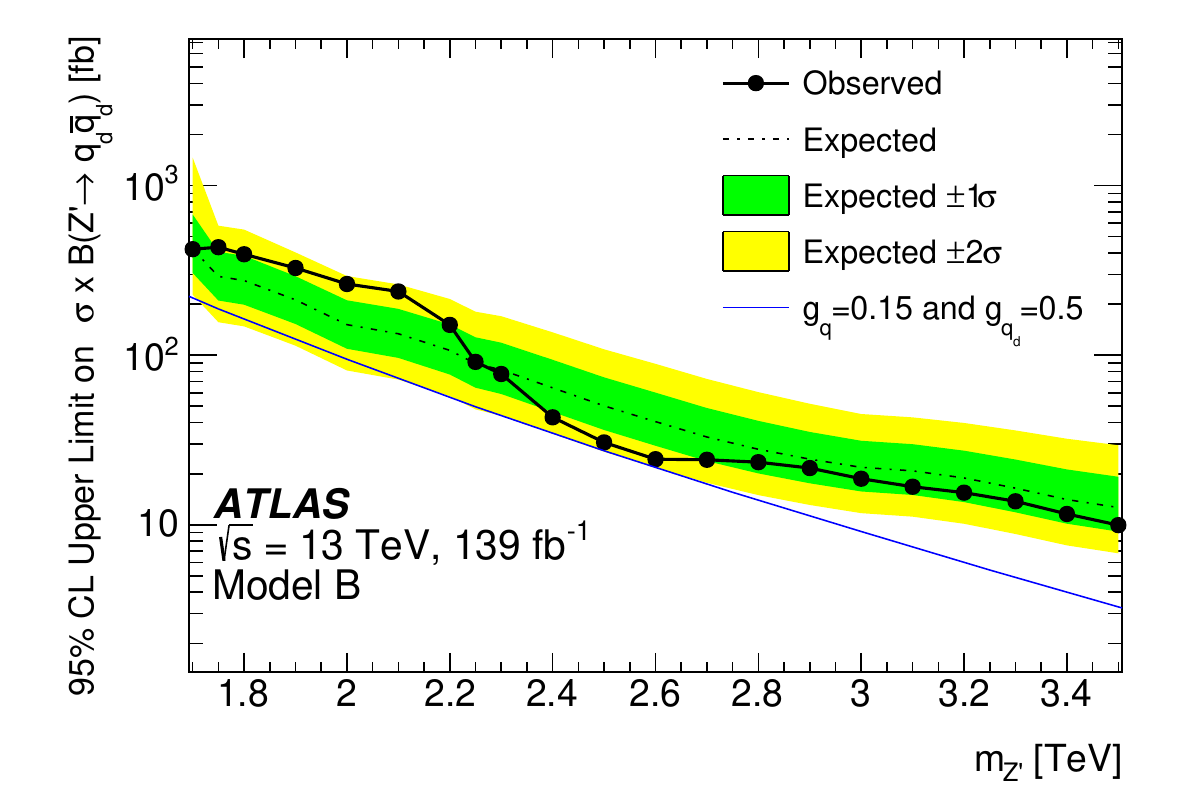}\label{fig:limits_modelB}} \\
\subfigure[]{\includegraphics[width=0.45\linewidth]{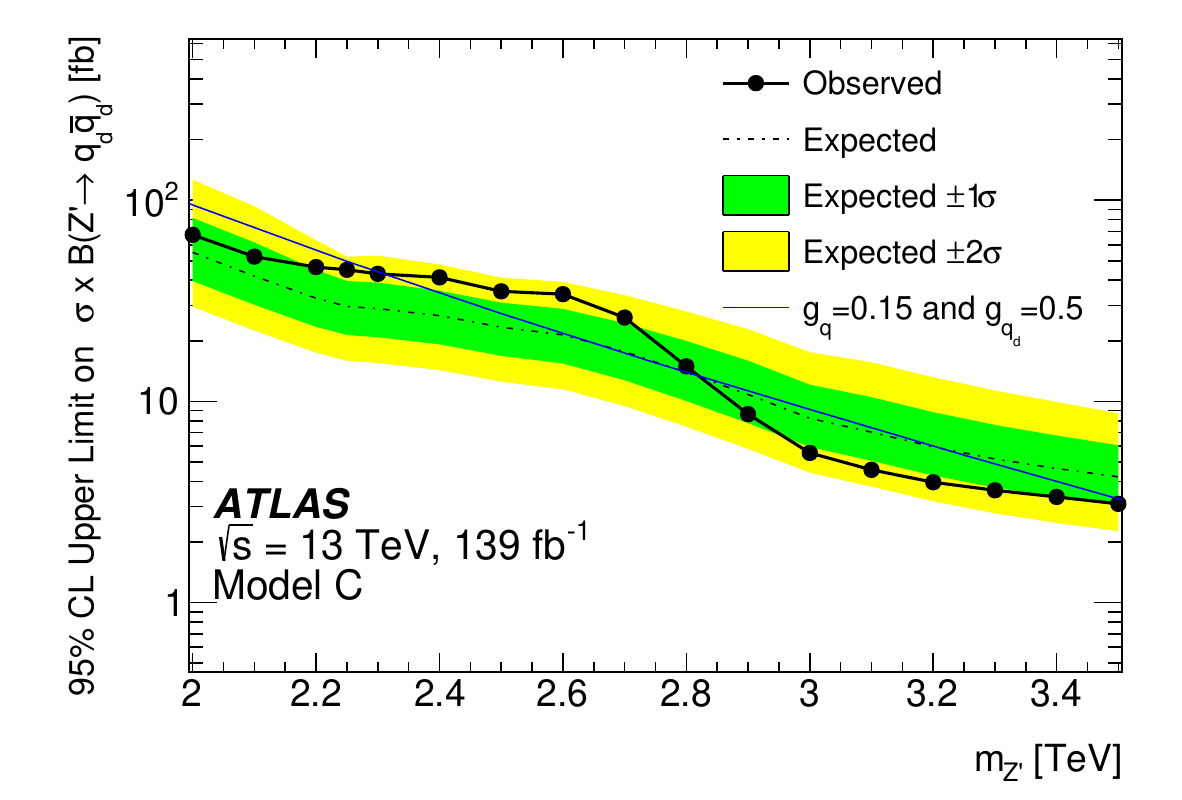}\label{fig:limits_modelC}}
\subfigure[]{\includegraphics[width=0.45\linewidth]{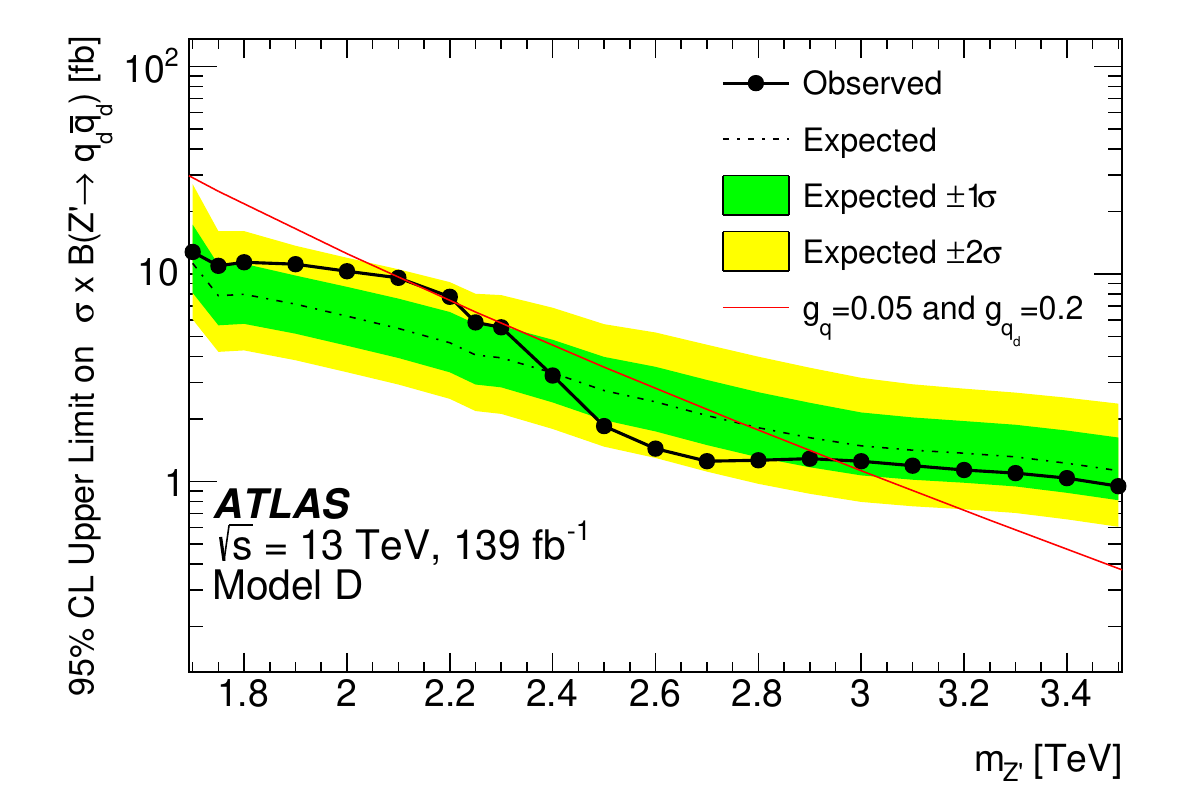}\label{fig:limits_modelD}} \\
\caption{Observed and expected limits at 95\% CL on the cross-section times branching ratio for the production of dark quarks through a \Zprime as a function of the \Zprime mass for models A, B, C and D (see the text for details). The darker and lighter shaded bands around the expected limits represent the $\pm1\sigma$ and $\pm2\sigma$ uncertainty range, respectively. The predicted cross-section times branching ratio is also shown as a thin solid line for $g_q=0.05$ and $g_{q_d}=0.2$ for models A and D, while values of $g_q=0.15$ and $g_{q_d}=0.5$ are used for models B and C.\label{fig:limits}}
\end{center}
\end{figure}

\section{Conclusion}
\label{sec:conclusion}
 
A search for a new resonance decaying into a pair of dark quarks was performed using 139~\ifb~of $\sqrt{s}=13$~\TeV~proton--proton collision data recorded with the ATLAS detector at the Large Hadron Collider between 2015 and 2018.
The invariant mass spectra of the two highest-transverse-momentum large-$R$ jets is probed after selecting jets with a high track multiplicity using a variable which decorrelates this selection from the \mjj distribution.
No significant excess is observed above the Standard Model multijet background whose shape is obtained using a data-driven method from a low-track multiplicity control region.
Constraints on the production cross-section times branching ratio to dark quarks are presented for four different dark QCD scenarios as a function of the \Zprime mass, excluding masses of up to 3.0~\TeV~for $g_q=0.05$ and $g_{q_d}=0.2$, depending on the model.

\section*{Acknowledgements}


We thank CERN for the very successful operation of the LHC, as well as the
support staff from our institutions without whom ATLAS could not be
operated efficiently.
 
We acknowledge the support of
ANPCyT, Argentina;
YerPhI, Armenia;
ARC, Australia;
BMWFW and FWF, Austria;
ANAS, Azerbaijan;
CNPq and FAPESP, Brazil;
NSERC, NRC and CFI, Canada;
CERN;
ANID, Chile;
CAS, MOST and NSFC, China;
Minciencias, Colombia;
MEYS CR, Czech Republic;
DNRF and DNSRC, Denmark;
IN2P3-CNRS and CEA-DRF/IRFU, France;
SRNSFG, Georgia;
BMBF, HGF and MPG, Germany;
GSRI, Greece;
RGC and Hong Kong SAR, China;
ISF and Benoziyo Center, Israel;
INFN, Italy;
MEXT and JSPS, Japan;
CNRST, Morocco;
NWO, Netherlands;
RCN, Norway;
MEiN, Poland;
FCT, Portugal;
MNE/IFA, Romania;
MESTD, Serbia;
MSSR, Slovakia;
ARRS and MIZ\v{S}, Slovenia;
DSI/NRF, South Africa;
MICINN, Spain;
SRC and Wallenberg Foundation, Sweden;
SERI, SNSF and Cantons of Bern and Geneva, Switzerland;
MOST, Taipei;
TENMAK, T\"urkiye;
STFC, United Kingdom;
DOE and NSF, United States of America.
In addition, individual groups and members have received support from
BCKDF, CANARIE, CRC and DRAC, Canada;
PRIMUS 21/SCI/017 and UNCE SCI/013, Czech Republic;
COST, ERC, ERDF, Horizon 2020, ICSC-NextGenerationEU and Marie Sk{\l}odowska-Curie Actions, European Union;
Investissements d'Avenir Labex, Investissements d'Avenir Idex and ANR, France;
DFG and AvH Foundation, Germany;
Herakleitos, Thales and Aristeia programmes co-financed by EU-ESF and the Greek NSRF, Greece;
BSF-NSF and MINERVA, Israel;
Norwegian Financial Mechanism 2014-2021, Norway;
NCN and NAWA, Poland;
La Caixa Banking Foundation, CERCA Programme Generalitat de Catalunya and PROMETEO and GenT Programmes Generalitat Valenciana, Spain;
G\"{o}ran Gustafssons Stiftelse, Sweden;
The Royal Society and Leverhulme Trust, United Kingdom.
 
The crucial computing support from all WLCG partners is acknowledged gratefully, in particular from CERN, the ATLAS Tier-1 facilities at TRIUMF/SFU (Canada), NDGF (Denmark, Norway, Sweden), CC-IN2P3 (France), KIT/GridKA (Germany), INFN-CNAF (Italy), NL-T1 (Netherlands), PIC (Spain), RAL (UK) and BNL (USA), the Tier-2 facilities worldwide and large non-WLCG resource providers. Major contributors of computing resources are listed in Ref.~\cite{ATL-SOFT-PUB-2023-001}.


\printbibliography
 
\clearpage
 
\begin{flushleft}
\hypersetup{urlcolor=black}
{\Large The ATLAS Collaboration}

\bigskip

\AtlasOrcid[0000-0002-6665-4934]{G.~Aad}$^\textrm{\scriptsize 102}$,
\AtlasOrcid[0000-0002-5888-2734]{B.~Abbott}$^\textrm{\scriptsize 120}$,
\AtlasOrcid[0000-0002-1002-1652]{K.~Abeling}$^\textrm{\scriptsize 55}$,
\AtlasOrcid[0000-0001-5763-2760]{N.J.~Abicht}$^\textrm{\scriptsize 49}$,
\AtlasOrcid[0000-0002-8496-9294]{S.H.~Abidi}$^\textrm{\scriptsize 29}$,
\AtlasOrcid[0000-0002-9987-2292]{A.~Aboulhorma}$^\textrm{\scriptsize 35e}$,
\AtlasOrcid[0000-0001-5329-6640]{H.~Abramowicz}$^\textrm{\scriptsize 151}$,
\AtlasOrcid[0000-0002-1599-2896]{H.~Abreu}$^\textrm{\scriptsize 150}$,
\AtlasOrcid[0000-0003-0403-3697]{Y.~Abulaiti}$^\textrm{\scriptsize 117}$,
\AtlasOrcid[0000-0002-8588-9157]{B.S.~Acharya}$^\textrm{\scriptsize 69a,69b,m}$
% ...
\bigskip
\\

% ...
$^{29}$Physics Department, Brookhaven National Laboratory, Upton NY; United States of America.\\
% ...
$^{35}$$^{(a)}$Facult\'e des Sciences Ain Chock, R\'eseau Universitaire de Physique des Hautes Energies - Universit\'e Hassan II, Casablanca;$^{(b)}$Facult\'{e} des Sciences, Universit\'{e} Ibn-Tofail, K\'{e}nitra;$^{(c)}$Facult\'e des Sciences Semlalia, Universit\'e Cadi Ayyad, LPHEA-Marrakech;$^{(d)}$LPMR, Facult\'e des Sciences, Universit\'e Mohamed Premier, Oujda;$^{(e)}$Facult\'e des sciences, Universit\'e Mohammed V, Rabat;$^{(f)}$Institute of Applied Physics, Mohammed VI Polytechnic University, Ben Guerir; Morocco.\\
% ...
$^{49}$Fakult\"{a}t Physik , Technische Universit{\"a}t Dortmund, Dortmund; Germany.\\
% ...
$^{55}$II. Physikalisches Institut, Georg-August-Universit\"{a}t G\"ottingen, G\"ottingen; Germany.\\
% ...
$^{69}$$^{(a)}$INFN Gruppo Collegato di Udine, Sezione di Trieste, Udine;$^{(b)}$ICTP, Trieste;$^{(c)}$Dipartimento Politecnico di Ingegneria e Architettura, Universit\`a di Udine, Udine; Italy.\\
% ...
$^{102}$CPPM, Aix-Marseille Universit\'e, CNRS/IN2P3, Marseille; France.\\
% ...
$^{117}$Department of Physics, New York University, New York NY; United States of America.\\
% ...
$^{120}$Homer L. Dodge Department of Physics and Astronomy, University of Oklahoma, Norman OK; United States of America.\\
% ...
$^{150}$Department of Physics, Technion, Israel Institute of Technology, Haifa; Israel.\\
$^{151}$Raymond and Beverly Sackler School of Physics and Astronomy, Tel Aviv University, Tel Aviv; Israel.\\
% ...
$^{a}$ Also Affiliated with an institute covered by a cooperation agreement with CERN.\\
$^{b}$ Also at An-Najah National University, Nablus; Palestine.\\
$^{c}$ Also at Borough of Manhattan Community College, City University of New York, New York NY; United States of America.\\
$^{d}$ Also at Center for High Energy Physics, Peking University; China.\\
$^{e}$ Also at Center for Interdisciplinary Research and Innovation (CIRI-AUTH), Thessaloniki; Greece.\\
$^{f}$ Also at Centro Studi e Ricerche Enrico Fermi; Italy.\\
% ...
$^{m}$ Also at Department of Physics, King's College London, London; United Kingdom.\\
% ...

\end{flushleft}


\end{document}